# Ultrafast and Bright Quantum Emitters from the Cavity Coupled Single Perovskite Nanocrystals

*Seongmoon Jun†, Joonyun Kim†, Minho Choi, Byungsu Kim, Jinu Park, Daehan Kim, Byungha Shin\*, Yong-Hoon Cho\**

†These authors contributed equally.

*Seongmoon Jun, Minho Choi, Byungsu Kim, Yong-Hoon Cho* – Department of Physics and KI for the NanoCentury, Korea Advanced Institute of Science and Technology (KAIST), 291 Daehak-ro, Yuseong-gu, Daejeon 34141, Republic of Korea

*Joonyun Kim, Jinu Park, Daehan Kim, Byungha Shin* – Department of Material Science and Engineering, Korea Advanced Institute of Science and Technology (KAIST), 291 Daehak-ro, Yuseong-gu, Daejeon 34141, Republic of Korea

ABSTRACT

Perovskite nanocrystals (NCs) have attracted increasing interest for the realization of single-photon emitters, owing to their ease of chemical synthesis, wide spectral tunability, fast recombination rate, scalability, and high quantum yield. However, the integration of a single perovskite NC into a photonic structure is yet to be accomplished. We successfully coupled a highly stable individual zwitterionic ligand-based $CsPbBr_3$ perovskite NC with a circular Bragg grating (CBG). The far-field radiation pattern of the NC inside the CBG exhibits high directionality toward a low azimuthal angle, which is consistent with the simulation results. We observed a 5.4-fold enhancement in brightness due to an increase in collection efficiency. Moreover, we achieved a 1.95-fold increase in the recombination rate. This study offers ultrafast (< 100 ps) single-photon emission and an improved brightness of perovskite NCs, which are critical factors for practical quantum optical applications.

Single-photon sources are fundamental elements in emerging quantum technologies such as quantum communication[1], quantum-enhanced metrology[2], and optical quantum computing[3]. Various systems have been investigated to establish single-photon sources, including spontaneous parametric down conversion[4], epitaxially-grown quantum dots[5], defects in bulk materials[6], and defects in two-dimensional materials[7]. Among these, inorganic cesium lead halide $CsPbX_3$ perovskite nanocrystals (NCs) are one of the most promising systems for single-photon emission owing to their wide wavelength tunability[8] in the visible wavelength range, ease of synthesis, and the scalability from processability in solution. Recently, single-photon emission at room-temperautre[9] has been also demonstrated. In addition, the blinking effect which hinders the use of colloidal quantum dots, other famous solution-based quantum emitters[10], has been reduced in recent reports on perovskite NCs due to the reduction of defect traps[11,12]. Moreover, perovskite NCs exhibited a long coherence time of 80 ps, which displayed the possibility to obtain transform-limited single-photon emission[13]. Particularly, $CsPbBr_3$ NCs have an emission wavelength of approximately 530 nm, which is suitable for underwater quantum communication[14].

The brightness of quantum emitters is crucial factor in practical quantum optical applications, such as quantum communications. The brightness of quantum emitters is influenced by factors such as the collection efficiency at the objective, quantum yield, and spontaneous emission rate. Therefore, two approaches can be used to improve the brightness of quantum emitters. The first one is enhancing the spontaneous emission rate by utilizing photonic cavities such as photonic crystals[15], distributed Bragg reflectors[16], and plasmonic cavities[17], which is also important for realizing indistinguishable single-photon emissions. In this perspective, perovskite NCs stand out compared to other quantum emitters because of their fast emission rate, which can be attributed to the bright triplet character of the lowest-lying exciton ground state[18] and their large oscillator strength[11]. Therefore, perovskite NCs are advantageous to achieve ultrafast single-photon sources even by coupling with a cavity with a moderate Purcell factor.

The second approach is to improve the collection efficiency by integrating to photonic structures, including circular Bragg gratings (CBG)[19], microlenses[20], rocket structures[21], and horn structures[22]. In particular, the integration of perovskite NCs with photonic structures is essential because of the limited collection efficiency originated from the random directional emission of the NCs. To date, the coupling of ensemble perovskite NCs with nanobeam structure[23,24] or plasmonic crystal[25] has been reported. However, they dealt with only high-density perovskite NCs, and the single-photon signal was not reported. Therefore, the integration of an individual perovskite NC with a photonic structure to enhance the recombination rate or the collection efficiency of a single photon emitter has not yet been fulfilled.

In order to integrate a single NC to the photonic structure, highly diluted colloidal NC systems are required. However, the widely used ligands for stabilizing colloidal perovskite NCs, a pair of oleylamine (OAm) and oleic acid (OA), is easily detached from the surface of NCs during the purification step before the dilution[26]. The labile binding nature of the OAm/OA pairs on the surface also affects the colloidal stability during the dilution of colloidal NCs. Thus, a substantial decrease in the photoluminescence (PL) quantum yield of the NCs upon dilution has been observed in OAm/OA-capped $CsPbBr_3$ NCs[27]. To overcome this issue, zwitterionic ligand-based lecithin-capped $CsPbBr_3$ colloidal NCs with remarkable colloidal stability upon dilution were synthesized[27].

In this work, we realized the integration of zwitterionic ligand-capped $CsPbBr_3$ perovskite NCs with the CBG structure. CBG structures have attracted much attention because they offer enhancements of not only recombination rate but also collection efficiency. The CBG, also known as the bullseye structure, is composed of a number of periodic concentric rings that can improve the directionality towards the normal direction by prohibiting the in-plane propagation of light, thereby increasing the collection efficiency. In addition, CBG structures can provide moderate Purcell enhancement over a wide spectral bandwidth owing to their relatively small quality factor (Q). We designed the CBG structure using the finite difference time domain (FDTD) method for a wavelength of 530 nm, which is the emission wavelength of the $CsPbBr_3$ NCs at a low temperature (10 K). We corroborated the Purcell effect and enhancement of collection efficiency at a wavelength of 530 nm through FDTD simulation. Subsequently, we fabricated the silicon nitride ($Si_3N_4$) CBG structure based on the designed dimensions. To investigate the influence of the CBG, the emissions from NCs placed inside and outside the CBG were compared. We measured the far-field irradiation patterns of the NCs inside and outside the CBG. The NC inside the CBG exhibited high directional emission to the normal direction, and the experimental results were well suited to the simulation data. Using a power-dependent PL experiment, we obtained the saturated intensity of the NCs to analyze the collection efficiency. The collection efficiency was improved by a factor of 5.4. Additionally, the time-resolved PL experiment was conducted to confirm the enhancement of the recombination rate. The radiative recombination rate was enhanced by a factor of 1.95. Furthermore, the second-order correlation was measured using the Hanbury-Brown and Twiss (HBT) experiment to affirm a single-photon emission. The corrected $g^{(2)}(0)$ value of 0.05 implies that the emission from NC is the quantum light. As a result, perovskite NC-based nano photonic structures showed ultrafast and bright emissions characteristics, which are critical properties for quantum optical applications. These methods of integration with the cavity can be also useful for realizing the other optical devices such as lasers and light-emitting diodes.

**Results**

**Sample description**

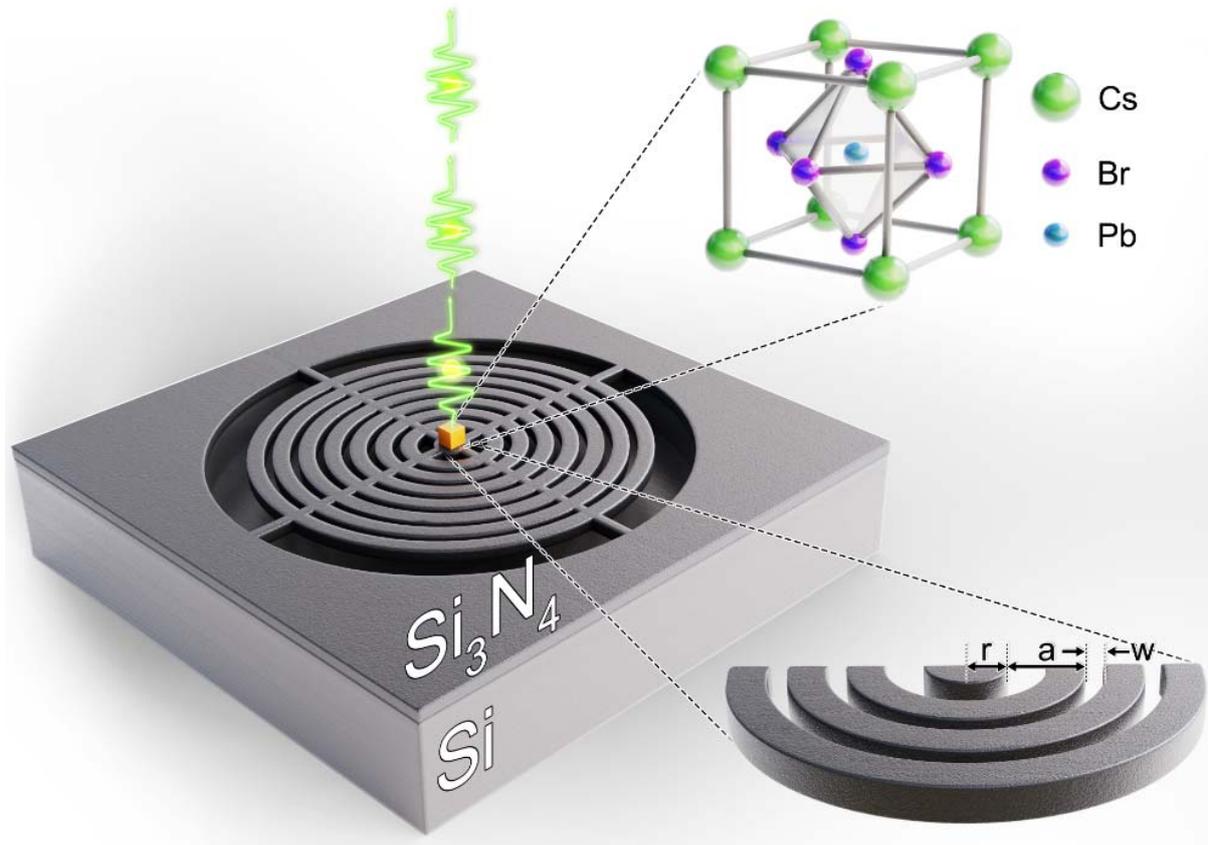

**Fig. 1 | Schematic of the experimental concept.** Illustration of the single-photon emission from the single $CsPbBr_3$ NC placed on the CBG. The CBG is made by the $Si_3N_4$ on silicon. Three parameters (r, a, w) are controlled to design.

Figure 1 shows the schematic of the integration of the CBG and a single perovskite NC. The material used for realizing the CBG is $Si_3N_4$ which provides a wide transparency window in the visible to near-infrared wavelength range. Furthermore, it has silicon compatibility, which is available with highly developed complementary metal-oxide-semiconductor fabrication infrastructures. To design the dimensions of the CBG, we used the parameters of the center disk radius (r), the trench width between the rings (w), and period length of the disks (a). The number of disks and their thickness were selected as 8 (Fig. S1) and 200 nm, respectively. These three variables were adjusted to optimize the Purcell factor at 530 nm using FDTD calculations. The optimized dimensions of the CBG are r = 312 nm, w = 100 nm, and a = 471 nm. In accordance with the calculation, $Si_3N_4$ CBGs were fabricated, and the perovskite NCs were spin-coated onto the $Si_3N_4$ CBG.

**Optical characterization of bare perovskite nanocrystal**

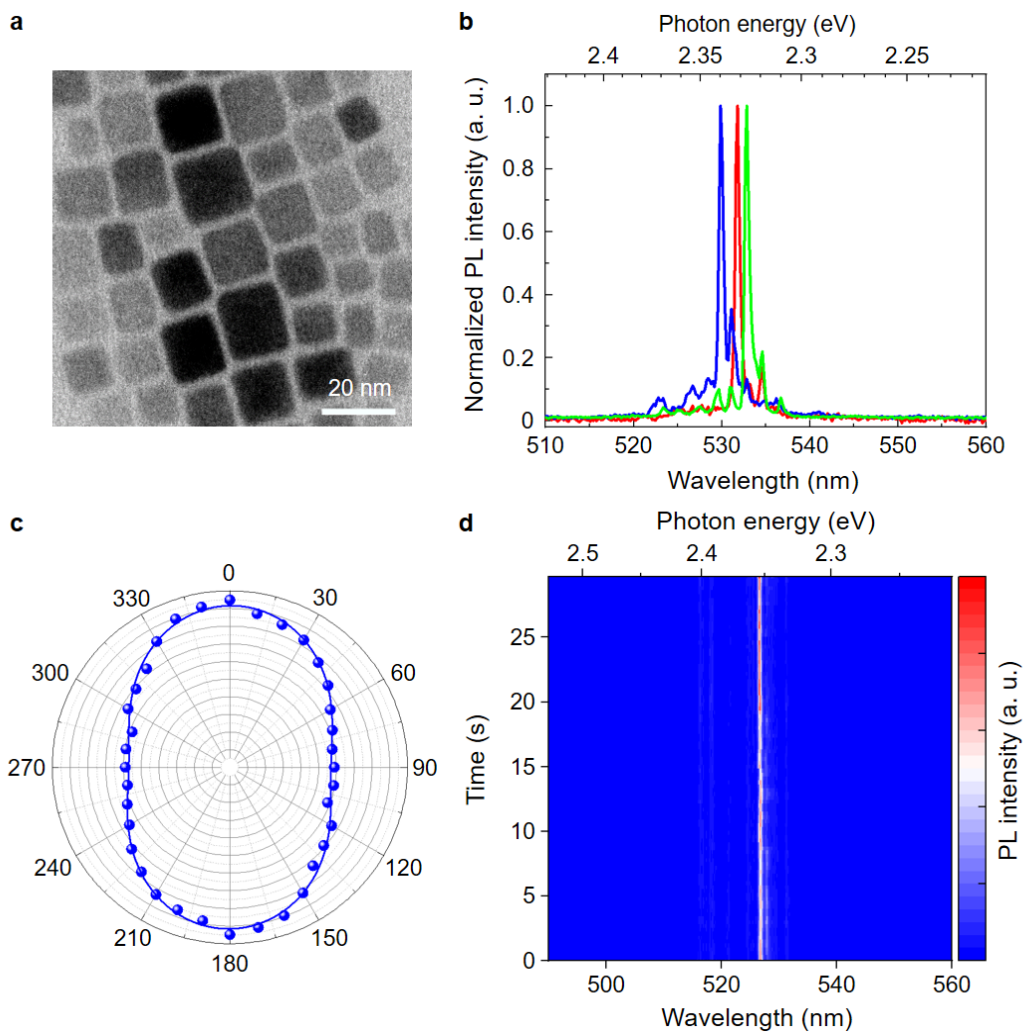

**Fig. 2 | Optical characterizations of single bare perovskite NCs at 10 K. a** TEM image of the zwitterionic ligand capped $CsPbBr_3$ NCs. The scale bar is 20 nm. **b** PL measurements of the three different single bare perovskite NCs on the $SiO_2$ substrate without CBG structure at 10 K. Sharp peaks are observed at a wavelength of around 530 nm. **c** Polarization-dependent PL result of the single bare perovskite NC. The perovskite NC shows a linear polarization. The

blue line is the fitting result. **d** PL measurement of the single bare perovskite NC for 30 s. The binning time is 300 ms. Blinking effect with dark states are not observed.

Figure 2a shows a transmission electron microscope (TEM) image of the zwitterionic ligand-capped $CsPbBr_3$ NCs. Zwitterionic ligand-based $CsPbBr_3$ NCs showed higher stability against the purification as well as upon dilution of the colloidal solution, which is a desirable feature for single-photon emitters compared to conventional OAm/OA ligands (Figs. S2, S3 and Supplementary Note 1). Cubic-shaped NCs, around 10 nm in size, were observed. To investigate the optical properties of the perovskite NCs before integrating with the CBG, a solution of perovskite NCs was spin-coated onto a $SiO_2$ substrate and the PL spectra were observed. Figure 2b shows the micro-PL results for the three single bare NCs without CBG at 10 K. Sharp peaks were observed which were distributed near 530 nm. Figure 2c represents the polarization-resolved PL emission of a single perovskite NC. For polarization-resolved PL, we measured the transmission intensity of the PL to the linear polarizer as a function of linear polarization angle, by adjusting the polarization angle of emission using a half-wave plate. The linear polarization from the NC indicates that the PL originates from the exciton because the emission from the trion is circularly polarized[18]. Since the exciton emission from spin-coated NCs has randomly directed linear polarization, the CBG, which does not have polarization selectivity due to the cylindrical symmetry, is a proper structure to couple with the NC. The PL emission was measured for 30 seconds to investigate the existence of the blinking effect, as shown in Fig. 2d. Although an insignificant fluctuation existed in the peak intensity, the blinking effect associated with dark states was not observed (Fig. S4).

**Simulation and fabrication of the circular Bragg grating**

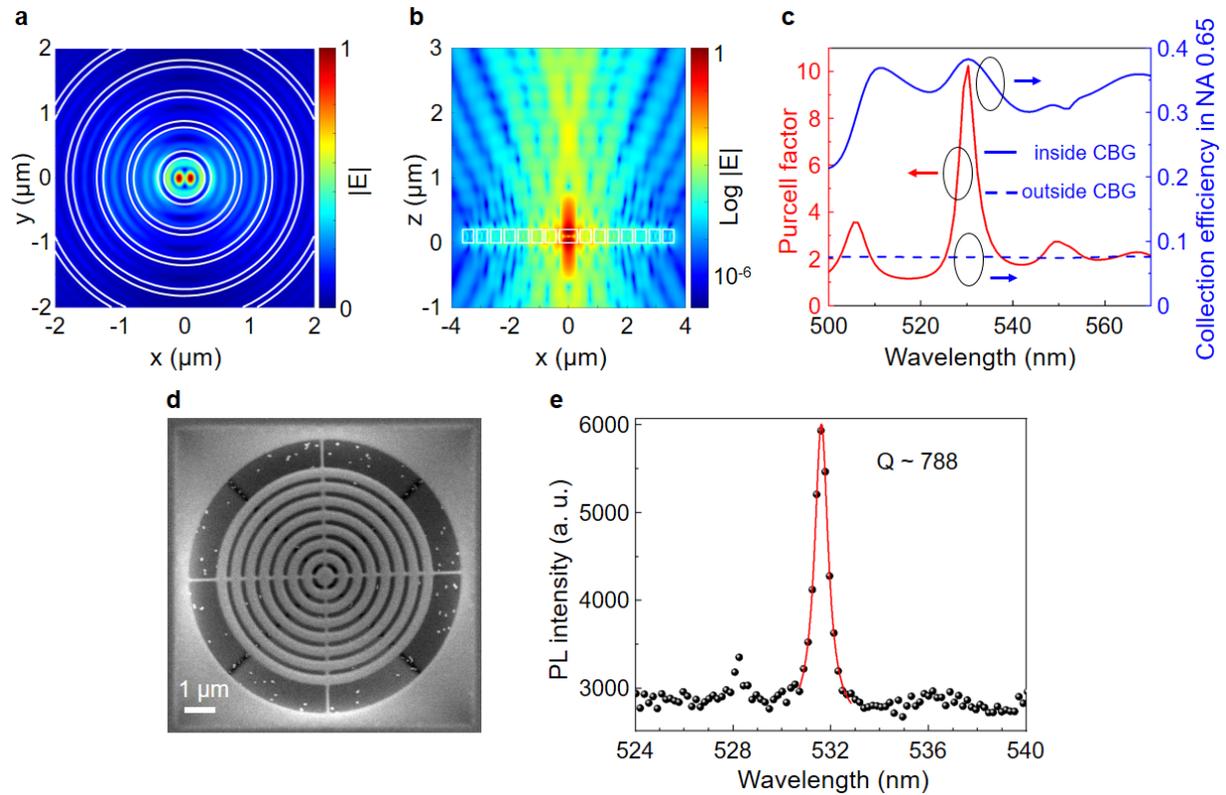

**Fig. 3 | Simulation and optical characterization of the CBG. a** Electric field magnitude distribution at a wavelength of 530 nm in the X-Y plane. **b** Log plot of electric field magnitude distribution at 530 nm in the X-Z plane. **c** Simulated Purcell factor (red, left axis) and collection efficiency as a function of wavelength. Collection efficiency is calculated on an objective basis with NA 0.65 in the case of a dipole located inside the CBG (solid blue, right axis) and outside the CBG (dashed blue, right axis). **d** SEM image of the CBG structure. The scale bar is 1 μm. **e** PL spectrum of the bare CBG with intrinsic $Si_3N_4$ emission. CBG mode is shown at a wavelength near 530 nm. The red line is the Lorentzian fitting graph of the mode peak.

To investigate the optical properties of the CBG, we calculated the electric field distributions by locating a single dipole on the center of CBG. Figure 3a shows the electric field in the X-Y plane. The difference of the refractive index at the interface of silicon nitride and air causes a reflection in the in-plane propagating light, thereby inducing a strong electric field at the center of the CBG. Therefore, the electric field is confined at the center of the CBG. Also, the electric field distribution in the X-Z plane is visualized in Fig. 3b. The electric field has high directionality to the out-of-plane direction owing to the inhibition of in-plane light propagation. Figure 3c represents the calculated Purcell factor and collection efficiency as a function of wavelength. The collection efficiency was calculated based on the objective lens used in the experiment with a numerical aperture (NA) of 0.65. In real applications, the NA value of the collective objective lens or optical fiber is limited, thus the high directional emission to the finite NA is important for enhancing collection efficiency. The Purcell factor is approximately 10 at a wavelength of 530 nm. In addition, the collection efficiency of the emitted light increased from 7.54 % outside the CBG to 38.3 % inside the CBG. If we utilize a metal layer[28] or distributed Bragg reflector as the bottom layer, the collection efficiency can be approximately doubled, owing to the reflection of the downward emission. It is worth noting that though the Purcell effect is realized in a narrow wavelength range, the improvement in collection efficiency is effective in a relatively broad wavelength range.

After designing the dimensions of the CBG, the CBG structures were fabricated using electron-beam lithography and inductive coupled plasma-reactive ion etching. Subsequently, the silicon was undercut by potassium hydroxide wet etching to create free standing CBG structures. Figure 3d shows the scanning electron microscopy (SEM) image of the fabricated CBG structure. To verify the cavity mode of the CBG, we measured the PL spectrum inside the CBG, as shown in Fig. 3e. Broad intrinsic emission from defects in the $Si_3N_4$ was used to probe the cavity mode. A mode peak near 530 nm was observed by measuring the PL emission from the CBG structure. The Q factor of the measured cavity mode was 788. Also, as shown in the mapping result with intrinsic $Si_3N_4$ PL (Fig. S5), the PL intensity of $Si_3N_4$ inside the CBG was enhanced compared to that of the $Si_3N_4$ emission from outside the CBG.

Enhancement of the intensity of intrinsic $Si_3N_4$ PL originates from the increase in collection efficiency.

**Far-field pattern**

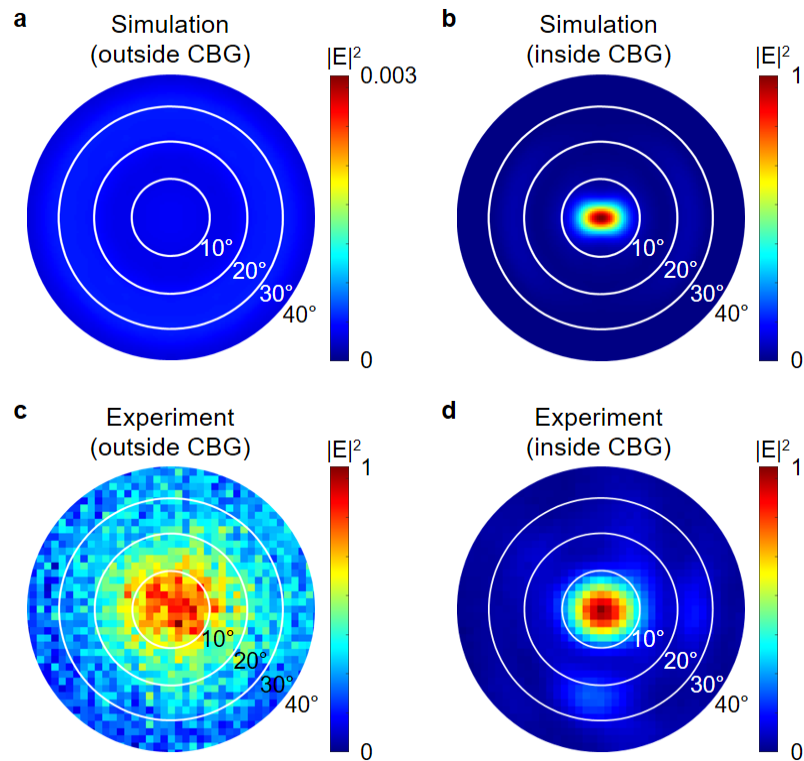

**Fig. 4 | Far-field pattern by simulations and optical measurements.** Simulated far-field radiation patterns expressed up to an angle of 40° from the dipole located **a** outside the CBG and **b** inside the CBG. The maximum intensity in the graph outside the CBG is 0.003, which is normalized to the maximum intensity in the graph inside the CBG. Measured far-field radiation patterns expressed up to an angle of 40° from the perovskite NC **c** outside the CBG and **d** inside the CBG. The graphs c and d were normalized by the maximum intensity of each result.

To measure the effect of the CBG structure on the emission directionality of the emitters, we calculated and measured the far-field radiation pattern from the single perovskite NCs inside and outside the CBG. Figures 4a and 4b show the simulated far field patterns from a single perovskite NC outside and inside the CBG, respectively. Because the NA 0.65 objective lens could collect emissions up to an angle of 40.54°, we visualized the far-field pattern up to 40°. The far-field emissions from the NC outside the CBG spread out in various angles. In contrast, we ensured that the far-field emission pattern from the NC inside the CBG exhibited high directionality to the angle lower than 10°, as depicted in Fig. 4b. Subsequently, micro-PL experiments were performed to measure the far-field patterns. To couple the single perovskite NC and CBG, we deposited 3 nm of $Al_2O_3$ and the diluted perovskite NCs were spin-coated onto the fabricated CBG structures. The sample was loaded onto a low-vibration cryostat for low-temperature measurements. In addition, a three-axis piezoelectric stage was utilized for high-precision movement of the sample location. A 405 nm pulsed laser generated by the second-harmonic generation of a wavelength-tunable Ti:Sapphire wavelength-tunable pulsed laser was exploited as the excitation laser. The PL spectra of the single perovskite NC inside the CBG were obtained by monitoring an image of the sample. Also, perovskite NCs outside the CBG were measured for comparison. Figures 4c and 4d illustrate the experimental far-field emission results from the single perovskite NC outside and inside the CBG, respectively. The experimental result for the far-field radiation pattern of the NC inside the CBG (spectrum is shown in Fig. S6) showed highly directional emission into low angles. This result matched well with the calculation result. Therefore, we can consider the measured far-field pattern as evidence of the coupling between the perovskite NC and the CBG structure.

**Enhancement of collection efficiency and lifetime**

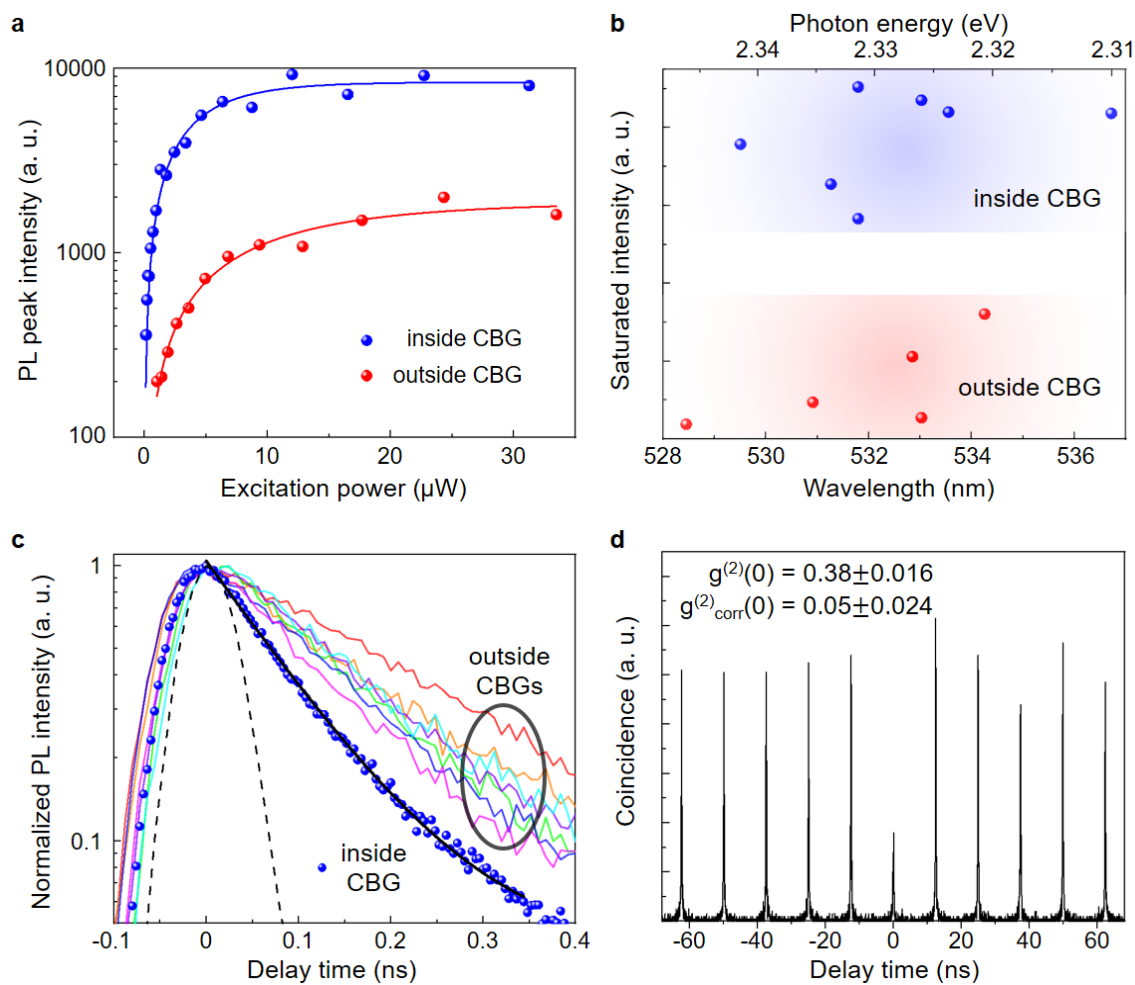

**Fig. 5 | Characterizations of the single NCs inside and outside the CBG. a** Power-dependent PL results of the single perovskite NC inside the CBG (blue circle) and outside the CBG (red circle). The red and blue lines represent the fitting curves. **b** Saturated intensities and wavelengths of the many single NCs inside the CBG (blue circle) and outside the CBG (red circle). A 5.4-fold enhancement of average saturated intensities was achieved. **c** Time-resolved PL results of the perovskite NC inside the CBG (blue dot) and NCs outside the CBG (colored solid line). The blue line is the mono-exponential fitting graph of the NC inside the CBG. The black dashed line is the instrumental response function. **d** Second-order correlation function of the emission from single perovskite NC inside the CBG. The measured $g^{(2)}(0)$ value is 0.38 and the background-corrected $g^{(2)}_{corr}(0)$ value is 0.05.

To study the enhancement of collection efficiency and lifetime, we performed power-dependent PL and time-resolved PL measurements at 10 K, respectively. Figure 5a shows the power-dependent PL intensities of the single perovskite NCs inside (identical NC with Fig. 4d, spectrum is shown in Fig. S6) and outside the CBG. The intensity for both NCs increased linearly with the excitation power in the low-power regime. However, saturation behavior was revealed at the high-power regime because a single two-level system cannot emit more than two photons simultaneously. The power-dependent intensity results were fitted using the following equation:

$$I = I_\infty \left(1 - e^{-\frac{P}{P_{Sat}}}\right),$$

where I is the intensity of the QD, $I_\infty$ is the saturated intensity of the QD, P is the pumping power of the laser, and $P_{sat}$ is the saturated pumping power. The saturated intensity was 8397 ± 335 and 1869 ± 148 at the NC inside and outside the CBG, respectively. To ensure that the increase in saturated intensity is also valid in other NCs, we performed power-dependent PL measurements for multiple NC emitters. Figure 5b represents the saturated intensities of several emitters inside and outside the CBG. The average saturated intensities of the emitters inside and outside the CBG were 7755 ± 1240 and 1439 ± 1201, respectively. Because we used pulsed laser excitation, the recombination rate of the emitters did not affect the saturated intensity. The measured saturated intensity is only differed by the collection efficiency of the emitters. Therefore, the increased saturated intensity of the NCs inside the CBG compared to that of the NCs outside the CBG originated from the coupling effect with CBG. As a result, it is worth noting that the saturated intensities of the NCs inside the CBG are enhanced by a factor of 5.4 as a consequence of raised collection efficiency. Furthermore, the enhancement of the saturated intensity is effective in all NCs inside the CBG. This implies that the collection efficiency can be enhanced regardless of the slight variation in the locations and emission wavelengths of the

NCs. Figure 5c shows the time-resolved emission decay of the NCs inside and outside the CBG. We measured the time decay of seven emitters outside the CBG. Each decay function was fitted using the mono-exponential function as following:

$$I = I_0 e^{-\frac{t}{\tau}} + y_0,$$

where I is the photon count as a function of time, $I_0$ is the maximum photon count, t is the delay time, $\tau$ is the lifetime, and $y_0$ is the background photon count. The average lifetime was 175.1 ± 54.6 ps for the emitters outside the CBG. In the case of the perovskite NC inside the CBG (identical NC in Fig. 4d, spectrum is shown in Fig. S6), the decay function was also fitted by the mono-exponential function with a lifetime of 89.9 ps (Fig. 5c). Only a few NCs inside the CBG were observed to exhibit a faster temporal decay than those outside the CBG. We deduced that not all NCs inside the CBG are influenced by the Purcell effect because of relatively narrow wavelength range of the Purcell effect (spectral mismatch) and the random distribution of the locations of the NCs (spatial mismatch). Even though, the measured NC inside the CBG in Fig. 5c shows the 1.95-fold enhancement in the recombination rate compared to the average of the NCs outside the CBG. To confirm that the emission from the NCs is the single-photon emission, we performed the HBT experiment. Figure 5d shows the second-order autocorrelation function of the emission from an identical NC with Fig. 4d. Because pulsed laser excitation was operated, the coincidence peaks were located with a periodicity of 12.5 ns, which is the time interval of the excitation laser. The $g^{(2)}(0)$ value was calculated by dividing the integrated coincidence count at zero delay by the average integrated counts of 10 peaks, which include five peaks located on either side of the zero delay. The $g^{(2)}(0)$ value of the measured NCs was 0.38 ± 0.016. In ideal single-photon sources, the $g^{(2)}(0)$ value becomes zero owing to the suppressed multi-photon events. However, the nonzero $g^{(2)}(0)$ value can be obtained due to uncorrelated emissions, such as the substrate background emission and other NCs near the measuring NC.

In our experiment, high $g^{(2)}(0)$ value may originate from the background emission of $Si_3N_4$. we can correct the $g^{(2)}(0)$ value with the following equation[29]:

$$1 - g^{(2)}(0) = \rho^2 \left(1 - g^{(2)}_{corr}(0)\right),$$

where $\rho$ is the signal to noise ratio, and $g^{(2)}_{corr}(0)$ is the corrected second-order correlation value. Therefore, by using the $\rho$ obtained from the spectrum (Fig. S6), the corrected $g^{(2)}_{corr}(0)$ is $0.05 \pm 0.024$. In fact, to exclude the effect from the substrate background emission, we conducted the HBT experiment of the perovskite NC on a silicon substrate and obtained a sufficiently lower measured $g^{(2)}(0)$ value (Fig. S7).

**Discussion**

We achieved clear antibunching behavior from the cavity coupled single perovskite NC, but the background emission from $Si_3N_4$ still causes the reduction of single-photon purity compared to the bare single perovskite NC on the silicon substrate (Fig. S7). Here, we can further reduce the $g^{(2)}(0)$ value by some techniques such as quasi-resonant excitation[30], resonant excitation[31], and nanoscale focus pinspot[32]. Meanwhile, the Purcell enhancement in this work was 1.95. This moderate Purcell enhancement may originate from the slight difference between the positions of the perovskite NC and the cavity mode, as the perovskite NCs are randomly distributed. This can be overcome by using nano-pockets[33] at the center of the CBG providing a deterministic integration between the NC and CBG. Nanoscale optical positioning[19] method can also improve the coupling efficiency between the NC and CBG. The undercut etching was used in this work to obtain a free standing CBG structure. However, the use of a metal layer[28] under the CBG could increase the reflection and lead to a higher collection efficiency of the emitters.

Integration with the CBG for perovskite NC quantum emitters can prove advantageous in scenarios where coupling to fiber optics[34] is desired because high directionality is crucial for effective coupling with optical fibers. It is worth mentioning that we were able to achieve ultrafast single-photon emission, which is under 100 ps. However, our perovskite NCs exhibited a slightly broadened linewidth, which is attributed to phonon-induced broadening or inhomogeneous broadening caused by spectral diffusion. Nevertheless, the Purcell enhancement provided by the CBG can be utilized to achieve transform-limited emissions if narrow-linewidth NCs can be synthesized. Additionally, for higher Purcell enhancement, the integration of perovskite NCs with other types of cavities such as photonic crystal and plasmonic bow-tie cavity[35] could be explored. Also, to improve the collection efficiency, the perovskite NCs could be coupled to a photonic structure such as a micro-lens or horn structure, in future work.

In summary, we successfully coupled a highly stable single zwitterionic ligand-based $CsPbBr_3$ perovskite NC with a CBG structure to realize ultrafast (< 100 ps) single-photon emission and 5.4-fold improved brightness. The significant enhancement of brightness will be helpful for perovskite NC to become reliable and valuable single-photon sources in quantum photonic technologies.

**Methods**

**Preparation of OAm/OA-capped CsPbBr$_3$ NCs**

The synthesis of OAm/OA-capped CsPbBr$_3$ nanocrystals was conducted by following previous report[26] with slight modification. For the purification, the crude solution was centrifuged at 10000 rpm for 5 min, and the precipitate was dispersed in 300 μL of hexane. After, the dispersion solution was centrifuged again at 10000 rpm for 3 min, and the precipitate was discarded. Additionally, 300 μL of hexane was added, and 1.2 mL of ethyl acetate was added, followed by centrifugation at 10000 rpm for 5 min. The precipitate was collected, and dispersed to the 0.5 mL of toluene. One more of purification step was applied, by addition of 1 mL of ethyl acetate to the colloidal solution, and the solution was centrifuged at 10000 rpm for 5 min. The precipitate was dispersed to the 0.5 mL of toluene.

**Preparation of zwitterionic ligand-capped CsPbBr$_3$ NCs**

The synthetic protocol of zwitterionic ligand-capped CsPbBr$_3$ nanocrystals was conducted by following previous report[27]. For the purification, acetone was added to the crude solution with 2:1 volumetric ratio, and the centrifugation was applied at 10000 rpm for 10 min. The precipitate was dispersed in 10 mL of toluene, and the acetone was mixed to the solution with 2:1 volumetric ratio and centrifuged again at 10000 rpm for 3 min. The purification step was repeated twice, and the final precipitate was dispersed in 2 mL of toluene.

**Numerical simulation**

FDTD numerical simulations are performed using commercial software (Ansys, Lumerical FDTD). The refractive index of Si$_3$N$_4$ is measured by ellipsometry and these measured values are used in the simulation. In the simulation, a single dipole is located on the 5 nm above the center of the CBG. The field monitor at the X-Y plane and X-Z plane and passes through the center of the CBG is used to record the electric field. By using the monitor at the upside of the CBG and conducting the near-to-far-field projection, the far-field patterns are obtained. The collection efficiency is computed as the ratio of the integration of the power in the far-field data to the total radiated power from a single dipole. The Purcell factor is calculated as the fraction between the radiated power of the dipole on the center of CBG and radiated power of the dipole in a vacuum.

**CBG fabrication**

CBG structures are fabricated using a 200 nm thick low-pressure chemical vapor deposited $Si_3N_4$ on silicon. The CBG patterns are written by electron-beam lithography with a positive e-beam resist. After that, 10 minutes of baking is performed at 140 °C. By inductive coupled plasma-reactive ion etching, the $Si_3N_4$ layer is etched following the e-beam lithography pattern. The e-beam resist is removed by acetone. Finally, the silicon etching for an undercut is conducted using potassium hydroxide in the condition of 80°C, for 10 minutes.

**TEM and SEM**

TEM images were collected using JEOL JEM-3010 microscope at 200 kV. SEM image was observed using FEI XL30.

**Single perovskite NC characterization**

We employed a wavelength-tunable femtosecond pulsed laser (Ti:Sapphire) with pulse duration of 200 fs and repetition rate of 80 MHz to serve as the excitation laser. The wavelength of the excitation laser was 405 nm, which is obtained by the second-harmonic generation of an 810 nm laser. To excite a single perovskite NC under low-temperature conditions, we used low-vibration closed-loop liquid-helium cryostat (Montana Instruments). Microscope objective lens (Mitutoyo; 50x; N.A., 0.65) was utilized to focus the laser on the perovskite NC and collect the emission of an identical NC. The collected PL spectra were analyzed using a high-resolution monochromator (Princeton Instruments, SP2750) and charge-coupled device detector. To measure the polarization-resolved PL, we manipulated the angle of the half-wave plate in front of the fixed linear polarizer. The far-field radiation patterns were obtained by measuring the momentum space of the emission. Here, spectral filtering by a bandpass filter was used. The time-resolved PL was observed by an avalanche photodiode (APD) (ID Quantique; ID100; temporal resolution, 40 ps) and a time-correlated single-photon counting system (Picoquant; Picoharp 300). We utilized the HBT experimental setup to measure the second-order autocorrelation, which included a 50:50 beam splitter and two APDs. In addition, two APDs were connected to a time-correlated single-photon counting system to obtain coincidence histograms that show the delay times between two detected photons at the APDs. We also applied spectral filtering with a bandwidth of approximately 0.7 nm in the second-order correlation measurement.

**Data Availability**

The data that support the findings of this study are available from the corresponding authors upon reasonable request.

**Acknowledgements**

This work was supported by the National Research Foundation (2022R1A2B5B03002560 and 2020M3E4A1080112) and Institute of Information & Communications Technology Planning & Evaluation (2020-0-00841) of the Korean government.

**Author contributions**

S.J., J.K., and M.C designed the experiment. S.J. and B.K. performed the optical characterizations. J.K., J.P., and B.S. synthesized the perovskites NCs. D. K. conducted the nanocrystal characterizations (TEM). S.J. and J.K. and Y.-H.C wrote the manuscript. Y.-H.C. conceived and supervised this project. All the authors discussed the results and commented on the manuscript.

**Competing interest**

The authors declare no competing interests.

**Additional information**

**Supplementary** is available

**Correspondence** and requests for materials should be addressed to B. S or Y.-H.C.